# Magnetic Generation and Switching of Topological Quantum Phases in a Trivial Semimetal α-$EuP_3$

Alex Hiro Mayo[1,2*], Hidefumi Takahashi[1,2,3], Mohammad Saeed Bahramy[4], Atsuro Nomoto[1], Hideaki Sakai[5], and Shintaro Ishiwata[1,2,3†]

[1] *Department of Applied Physics, University of Tokyo, Bunkyo-ku, Tokyo 113-8656, Japan*

[2] *Division of Materials Physics, Graduate School of Engineering Science, Osaka University, Toyonaka, Osaka 560-8531, Japan*

[3] *Center for Spintronics Research Network (CSRN), Graduate School of Engineering Science, Osaka University, Toyonaka, Osaka 560-8531, Japan*

[4] *Department of Physics and Astronomy, University of Manchester, Oxford Road, Manchester M13 9PY, United Kingdom*

[5] *Department of Physics, Graduate School of Science, Osaka University, Toyonaka, Osaka 560-0043, Japan*

## Abstract

Topological materials have drawn increasing attention owing to their rich quantum properties, as highlighted by a large intrinsic anomalous Hall effect (AHE) in Weyl and nodal-line semimetals. However, the practical applications for topological electronics have been hampered by the difficulty in the external control of the band topology. Here we demonstrate a magnetic-field-induced switching of band topology in α-$EuP_3$, a magnetic semimetal with a layered crystal structure derived from black phosphorus. When the magnetic field is applied perpendicular to the single mirror plane of the monoclinic structure, a giant AHE signal abruptly emerges at a certain threshold magnetization value, giving rise to a prominently large anomalous Hall angle of $|\Theta_{AHE}| \sim 20^\circ$. When the magnetic field is applied along the inter-layer direction, which breaks the mirror symmetry, the system shows a pronounced negative longitudinal magnetoresistance. On the basis of electronic structure calculations and symmetry considerations, these anomalous magneto-transport properties can be considered as manifestations of two distinct topological phases: topological nodal-line and Weyl semimetals, respectively. Notably, the nodal-line structure is composed of bands with the same spin character and spans a wide energy range around the Fermi level. These topological phases are stabilized via the exchange coupling between localized Eu-4*f* moments and mobile carriers conducting through the phosphorus layers. Our findings provide a realistic solution for external manipulation of band topology, enriching the functional aspects of topological materials.

*mayo@qm.mp.es.osaka-u.ac.jp

†ishiwata@mp.es.osaka-u.ac.jp

## I. INTRODUCTION

Topological semimetals (TSs) are a new class of quantum materials hosting non-trivial massless fermionic states, which are derived from linearly dispersing bands crossing at the Fermi level [1–8]. Such band crossings occur in the bulk and are formed at either isolated or continuous points in momentum space, the latter being commonly known as topological nodal-line semimetals. TSs are further categorized into two groups—Weyl and Dirac types—depending on whether they lack or hold spin-degeneracy throughout their Brillouin zone, respectively. Owing to their relativistic nature, such band crossings generate a local but large internal magnetic flux, known as the Berry curvature, that can give rise to intriguing transport phenomena such as the intrinsic anomalous Hall effect (AHE) [9–12]. By controlling the time and space inversion symmetries, one such topological phase can in principle be turned into another. Externally this can be achieved by applying a magnetic field or via mechanical distortion, enabling practical manners in which to exploit the band topology and embed it into devices with novel spintronic functionalities of relevance for future energy and information technologies.

Thus far, various TSs containing 3*d* and 4*f*-elements have been discovered in nature or grown under controlled conditions [4–8,13–20], incorporating magnetism as a possible controlling parameter. However, the expectations for the magnetic tunability of their electronic states and functionalities have not been met, owing to multiple factors. For example, the magnetic states in 3*d*-TSs are generally irresponsive to the magnetic field due to the strong chemical hybridization of 3*d* orbitals with the rest of the electronic structure. More critically, in most of these systems, the relevant topological nodes are distant from the Fermi level, $E_F$. As such, they are untraceable in transport measurements unless extensive band tuning is performed, which is not always possible, or at least leads to collateral damages that can affect the desired properties of the system. Therefore, significant effort is being dedicated towards the discovery of TSs with tunable topological nodes in immediate proximity to their Fermi level. In this paper, we introduce a magnetic semimetal, α-$EuP_3$, with a simple, yet profoundly tunable electronic structure that ideally meets these criteria.

## II. RESULTS

α-$EuP_3$ is a layered magnetic semimetal with a monoclinic crystal structure (space-group type *C*2/*m*) whose *b*-vector serves as a two-fold rotational axis normal to a mirror plane [21–23] (Figs. 1(a) and (b)). The bulk of α-$EuP_3$ is comprised of two-dimensional (2D) puckered polyanionic layers of phosphorus that stack along the *c*-axis, with $Eu^{2+}$ cations in between. The structure of the P-layers is closely related to that of orthorhombic black phosphorus [24–26] and can be derived by removing 1/4 of its P-centers. The remaining P-layers form a closed shell by receiving valence electrons from the 6*s*-orbitals of Eu cations, to the extent that the whole system gains a semi-metallic or semiconducting character with half-filled Eu-4*f* shells. This black-phosphorus-derived quasi-2D nature coordinated with localized *f*-moments is the key aspect of α-$EuP_3$ which enables large spin-polarization in the intrinsically non-magnetic P-layers via energetical proximity of their carriers with the immobile Eu-4*f* moments. The locality of the Eu-4*f* moments and their close proximity to $E_F$ further allows a directional control of the resulting spin-charge coupling that can be externally enforced by a magnetic field. Such highly field-controllable magnetic layers enable us to increasingly induce a large internal field by aligning the local moments, which, as will be shown shortly, yields a topologically non-trivial electronic structure that can similarly be controlled using a magnetic field.

To demonstrate the effect of Eu-4*f* moments on the electronic structure, Figure 1(c) shows the relativistic electronic structure of α-$EuP_3$ in the paramagnetic (PM) and field-induced ferromagnetic (FM) phases. The dashed lines are spin degenerate bands calculated by treating the 4*f* states as core levels, thus, represent the PM phase band structure under $B = 0$. Here, two consecutive band inversions occur at the time-reversal invariant momenta Z and Γ points of the Brillouin zone. The combination of these two band inversions keeps the $Z_2$-index of the occupied bands equal to zero, meaning that the system is topologically trivial under zero magnetic field in the PM phase (Fig. 1(g)). The colored solid lines are the corresponding energy bands when the Eu 4*f*-moments are fully aligned along the *c**-axis, thus, represent the field-induced FM phase. This makes the 4*f* orbitals of divalent Eu ions half-filled $4f^7$ in the highest possible magnetic state $S = 7/2$ (i.e., 7 $\mu_B$), due to the implications of the Hund's rule. Our calculations reveal that these filled *f*-bands lie ~ 1.5 eV below $E_F$. Such proximity to $E_F$ allows the localized *f*-moments to be coupled to each other through a carrier-mediated exchange interaction mechanism [27–29]. As a result of such interaction with itinerant carriers, the valence and conduction bands near $E_F$ undergo opposite spin splitting depending on how the carriers couple to the localized *f*-moments [30–32]. The valence bands, consisting of occupied P-*p* orbitals, show a Kondo-like interaction with an antiparallel coupling to the Eu-*f* moments, whereas the conduction bands, consisting of unoccupied Eu-*s* orbitals, undergo a Hund-like parallel coupling. Consequently, spin-up valence bands are forced to stay above their spin-down counterparts, whereas conduction bands follow an opposite spin splitting. Another important consequence of such spin splitting is the immunity of particular types of crossings between the conduction and valence bands, that could lead to the formation of a nodal line or Weyl nodes depending on whether the magnetic field breaks the mirror plane or respects it. These scenarios are schematically illustrated in Fig. 1(g) and will be discussed in more detail in the context of magneto-transport, after a description of the experimental data.

Figures 1(d)-(f) show temperature dependences of the magnetization, resistivity, and Seebeck coefficient, respectively. The resistivity demonstrates metallic behavior and a sharp peak at $T_N$ = 8.2 K, where it exhibits antiferromagnetic (presumably helimagnetic [33,34]) ordering. This peak is significantly suppressed by aligning the local spins with an external magnetic field, indicating the presence of a magnetic coupling between local spins and carriers, as expected from the band calculation. Despite the field dependence and anisotropy of the resistivity seen at higher temperatures, the corresponding Seebeck coefficients along the *a*- and *c**-axes show nearly identical and *T*-linear behavior below ~ 150 K, suggesting that the carrier types or density of states are hardly affected by direction or magnetic field in this temperature region.

Figures 2(a)-(f) show the magnetic field dependences of magnetization as well as the longitudinal and transverse (Hall) resistivities. The magnetization shows multiple magnetic-field-dependent transitions in the temperature range below $T_N$ (Figs. 2(a) and (d)). Notably, the magnetization saturates at low field (~ 2.5 T at 2 K) in both directions, showing an isotropic tunability of the Eu-4*f* local spins. In both $\boldsymbol{B} \parallel \boldsymbol{b}$ and $\boldsymbol{B} \parallel \boldsymbol{c}^*$ configurations at $T$ = 2 K, the longitudinal resistivity $\rho_{aa}$ shows a sharp anomaly along with a hysteresis corresponding to the magnetic transition (Figs. 2(b) and (e)), which reinforces the validity of a strong *f*-*p* coupling in this system. One notable difference between the two configurations is the behavior of the Hall resistivities $\rho_{c^*a}$ and $\rho_{ba}$ (Figs. 2(c) and (f)). While both measurements show similar linear behavior at $T$ = 150 K, a particularly large deviation from the normal Hall effect is observed in $\rho_{c^*a}$ at lower temperatures, suggesting a large additional Hall contribution with strong anisotropy. Another characteristic behavior can be seen in the magnetoresistance (MR). Figs. 2(g) and (h) show $B$ dependences of $\rho_{cc}$ under $\boldsymbol{I} \parallel \boldsymbol{B} \parallel \boldsymbol{c}$ measured at selected temperatures, and $\Delta\rho/\rho(0) = \{\rho(B) - \rho(0)\}/\rho(0)$ measured under three different configurations at $T$ = 15 K, respectively. Reflecting the presence of localized spins and spin-charge coupling, negative MR owing to the suppression of spin fluctuations is observed in all configurations. Here, it is noteworthy that the negative MR is pronounced for the configuration with $\boldsymbol{I} \parallel \boldsymbol{B} \parallel \boldsymbol{c}$, whereas it is nearly identical for the others as shown in Fig. 2(h). This characteristic will be further discussed in section III.

The Hall resistivity in magnetic systems is described by the empirical formula $\rho_H = \rho_H^N + \rho_H^A + \rho_H^T$. The first term is the normal Hall contribution $\rho_H^N = R_0B$ due to the Lorentz force, the second is the AHE, and the third is the topological Hall effect derived from spin textures associated with a topological number, such as skyrmions [35,36]. The topological Hall effect is unlikely to be the case for α-$EuP_3$ because the deviation from the normal Hall effect is persistent at $T > T_N$, where the system has no long-range spin ordering. This leads us to describe the Hall effect in α-$EuP_3$ as $\rho_H = \rho_H^N + \rho_H^A$. Further on, we will mainly focus our analysis on the $T > T_N$ region. In order to extract $\rho_H^A$, $\rho_H^N$ must be determined from experimental data, which is not necessarily definitive and thus requires a certain degree of ad hoc assumptions. Several approaches for extracting $\rho_H^A$ are presented in the Supplemental Material [37]. Since each approach gives reasonably similar results in the qualitative level, here we estimate $\rho_H^N$ by using $\rho_H^{T = 150\text{ K}}$, which is far above $T_N$ and approximately reflects the normal contribution to the Hall effect.

Since magnetization $M$ characterizes the proximity-induced exchange splitting, it is meaningful to investigate the relation between $M$ and the AHE. Figure 3(a) shows $\rho_{c^*a}{}^{A} = \rho_{c^*a} - \rho_{c^*a}{}^{T = 150\text{ K}}$ in the space of $M$ and $T$, showing a $T$-independent relationship between $M$ and $\rho_{c^*a}{}^{A}$ such that the anomaly clearly emerges at a certain threshold of $M \sim 1.8\ \mu_B$/Eu (= $M_C$). In other words, the trigger of the anomaly is the magnitude of $M$ rather than a specific spin texture, which strongly suggests that the band splitting plays a critical role in the emergence of the AHE. Figure 3(b) shows the $M$ dependence of the anomalous Hall conductivity $\sigma_{c^*a}{}^{A} = -\rho_{c^*a}{}^{A}/(\rho_{aa}\rho_{c^*c^*} - \rho_{ac^*}\rho_{c^*a})$. Interestingly, we see a clear deviation from the conventional $\propto M^1$ dependence as $M$ exceeds $M_C$. Although the $\sigma_{c^*a}{}^{A}$-$M$ curve shows a variation depending on how we estimate $\rho_H{}^{N}$, as mentioned above, the overall behavior reasonably follows the colored guide-to-eyes shown in Fig. 3(b) (see Fig. S9 [37]). We can also quantify the AHE using the anomalous Hall angle $\Theta_{AHE} = \tan^{-1}(\sigma_{c^*a}{}^{A}/\sigma_{aa})$. As presented in Table 1 and in Fig. S11 [37], the value of $|\Theta_{AHE}|$ in this work reaches 14° ~ 28° at low temperatures, which is notably large even when compared with the maximum values observed in previously reported materials [4,6,7,38–43] including topological candidates. These experimental results suggest an unconventional origin of the anomaly.

The AHE has been well studied in FM metals [9,44,45] and can be classified into the following three mechanisms. The first mechanism is known as the intrinsic contribution, related to $\boldsymbol{r}$ (real) space and $\boldsymbol{k}$ (momentum) space Berry curvature $\boldsymbol{b}(\boldsymbol{R})$ ($\boldsymbol{R} = \boldsymbol{r}, \boldsymbol{k}$). The former arises in non-coplanar spin textures with finite scalar spin chirality $\chi_{ijk} = \boldsymbol{S}_i \cdot (\boldsymbol{S}_j \times \boldsymbol{S}_k)$, where $\boldsymbol{S}_n$ are spins, which is again unlikely to be the case here. In contrast, the latter originates from the multi-bands in $\boldsymbol{k}$-space and is described as

$$\boldsymbol{b}(\boldsymbol{k}) = \nabla_{\boldsymbol{k}} \times \{\mathrm{i}\langle\varphi|\partial/\partial\boldsymbol{k}|\varphi\rangle\},$$

where $\varphi$ is the electronic wave function [11]. The second and third mechanisms can be described by the skew-scattering and side-jump contributions, which are both related to spin-orbit coupled impurity scattering. While the skew-scattering contribution can be ruled out because it only arises in highly conductive metals, the side-jump contribution can be present. However, since the atomic spin-orbit coupling term is expected to be generally small in the relevant bands [46], the side-jump contribution is also unlikely to dominate the large AHE. Moreover, these contributions are expected to scale with $M$, and it is unlikely that the side-jump scattering, caused by the impurity in the material, would abruptly enhance under monotonically increasing $M$. Such nature of the extrinsic components is incompatible with the observed unconventional $M$ dependence that turns on at $M \sim M_C$. Therefore, the most likely dominant origin of this feature is the intrinsic contribution derived from the $\boldsymbol{k}$-space Berry curvature. In this case, the direction of the Berry curvature would strongly depend on the band structure anisotropy in $\boldsymbol{k}$-space.

To verify this, Figure 4 shows the detailed electronic band structure of α-$EuP_3$ in various magnetic configurations. The PM phase ($B = 0$) is given in Fig. 4(a). Figures 4(b)-(d) and Figs. 4(e)-(g), in contrast, show the respective electronic structures when the Eu $f$-moments are forced to be fully FM ordered ($M = 7\ \mu_B$/Eu) along the $b$-axis (FM-b) and $c$*-axis (FM-c). Under the application of $B$, it can be reasonably assumed that the exchange splitting continuously increases as $M$ increases, giving rise to a continuous band evolution from Fig. 4(a) to Fig. 4(b) or (e). In FM-b, since the magnetic field is normal to the mirror plane, all the energy states maintain their mirror symmetry, even though they are no longer protected by the time-reversal symmetry. As a result, any band crossing between the conduction and the valence bands can be immune against hybridization by the mirror operators as long as they have the same spin characters. This requirement is ideally met at and in the vicinity of $E_F$ due to the opposite spin splitting of the conduction and valence bands, as discussed earlier. Since this crossing does not require any other symmetry operation, it can form a closed-loop nodal line within the mirror plane [2,47,48] (Fig. 4(c)). Interestingly, this nodal-ring around the Γ-point can span a large energy window, encompassing $E_F$ (Fig. 4(d)), and hence contribute directly to the anomalous Hall part of the transport. This energy spanning is due to the layered nature of α-$EuP_3$, which causes anisotropic electronic dispersions, ranging from steep ones for any $\boldsymbol{k}$-vector in the intra-layer direction, to nearly flat dispersions for the $\boldsymbol{k}$-vectors in the inter-layer direction. As shown in Fig. 4(b), and also in the schematic panel of Fig. 3(a), the energy difference between the top and bottom of the nodal-ring reaches ~ 250 meV, meaning that this material can naturally exhibit a giant AHE, emerging from such topological nodal line, in a wide range of chemical potentials (see Fig. S8 [37]). Thus, usual structural imperfections and disorders cause no serious complication. The fact that this phase can be easily stabilized using a relatively weak magnetic field applied along the crystalline *b*-axis further signifies the controllability aspect of such novel topological states in α-$EuP_3$. It is also worth noting that this is a unique experimental example of the realization and detection via quantum transport of such fully spin-polarized nodal-ring, i.e., a Weyl nodal-ring consisting of bands of the same spins.

The situation in FM-c is crucially different. Interestingly, the nodal-ring discussed earlier vanishes because the new alignment of Eu *f*-moments along the *c**-axis breaks the mirror symmetry. Instead, a single pair of Weyl nodes appear along the high-symmetry direction Γ-Y (Figs. 4(e), (f), and (g)). Since the direction of the *f*-moments are parallel to the mirror plane, bands of opposite spins are allowed to cross each other without hybridization as long as they share the same mirror eigenvalues [2]. This is the only way to avoid hybridization while respecting all existing symmetries.

## III. DISCUSSION

Let us discuss the origin of the anomalous transport based on the calculated band characteristics. The contribution of the Berry curvature on the Hall resistivity depends on the measurement configuration, as seen in the following relation [1]

$$\rho_{\alpha\beta}^{A} \propto \varepsilon_{\alpha\beta\gamma}\, b^{\gamma}(\boldsymbol{k}),$$

where $\varepsilon$ is the Levi-Civita symbol with $\{\alpha, \beta, \gamma\} = \{a, b, c^*\}$. When $\boldsymbol{B} \parallel \boldsymbol{c}^*$ (FM-c), the Weyl nodes emerge along Γ-Y $\parallel \boldsymbol{a}^*$, thus the Berry flux is expected to have projection components $b^{c^*}(\boldsymbol{k})$, along the *c**-axis contributing to $\rho_{c^*a}^{A}$ as seen in Fig. 2(f), as well as $b^{a}(\boldsymbol{k})$, along the *a*-axis mainly contributing to $\rho_{bc^*}^{A}$ (see Fig. S13 [37], measured under an unconventional Hall measurement configuration [$I \parallel B$]). For such a Weyl semimetal state, we should expect to observe a pronounced negative longitudinal MR caused by the chiral anomaly, in the configuration of $\boldsymbol{I} \parallel \boldsymbol{B}$ [1]. This is indeed demonstrated in Figs. 2(g) and (h). At high magnetic fields, the negative MR for $\rho_{cc}$ with $\boldsymbol{I} \parallel \boldsymbol{B} \parallel \boldsymbol{c}$ is notably enhanced even when compared with $\rho_{aa}$ with $\boldsymbol{I} \parallel \boldsymbol{B} \parallel \boldsymbol{a}$, where the Lorentz force is absent as well. This suggests an additional contribution of the Weyl nodes on the negative MR, other than the suppression of spin fluctuations (see also Fig. S15 [37]). Note that the positive MR forming a kink at low magnetic fields is found in a wide temperature range as shown in Fig. 2(g). Although this kink can be a characteristic feature of the magnetic generation of the Weyl states, further studies are necessary to reveal the origin of this behavior.

When $\boldsymbol{B} \parallel \boldsymbol{b}$ (FM-b), which is the case for Figs. 2(a)-(c), the *f*-moments are aligned normal to the mirror plane. In this measurement configuration, the mirror symmetry is preserved and nodal-rings emerge within the mirror plane. In the low $M$ region, the topological band crossings are wholly above $E_F$; thus, the $\boldsymbol{k}$-space Berry curvature hardly contributes to the Hall transport. At larger $M$ ($> M_C$), where the exchange splitting increases, the anisotropic nodal-ring around the Γ-point starts to partially cross $E_F$ in the Γ-Y direction, as depicted in the schematic panels of Fig. 3(a) and Fig. 4(d), leading to a $\boldsymbol{k}_{\Gamma X} \parallel \boldsymbol{b}$ projection of the Berry curvature $b^{b}(\boldsymbol{k})$, manifesting as the giant anomaly in Fig. 2(c). As for the prominently large $|\Theta_{AHE}|$, not only the large numerator $\sigma_H^{A}$ but also the small denominator $\sigma_{aa}$ is presumed to be an essential factor, which can be understood as a result of few trivial bands composing the Fermi surface rather than the dirtiness of the material (see Fig. S11 [37]). In other words, the combination of the topological nodes and the simple band manifold at $E_F$ enables the coexistence of a large $\sigma_H^{A}$ and small $\sigma_{aa}$ in α-$EuP_3$. Combining all the experimental and theoretical evidence, it is reasonable to presume that the topological nodes are generated under a sufficient net magnetic moment, strong enough to break or respect the mirror symmetry, and then can be tuned towards the Fermi level by increasing the degree of spin polarization of the localized *f*-moments.

Our experimental and theoretical work on α-$EuP_3$ presents a case for magnetic-field-triggered generation and switching of topological states, as illustrated in Fig. 1(g). The gigantic AHE in the nodal-line phase persists up to high magnetic fields, which is consistent with our theoretical expectation of the existence of tilted nodal-rings spanning a wide energy range around $E_F$. Crucially, these findings lead to a more general concept: that the combination of two critical components—a highly anisotropic crystal structure with a relevant mirror plane and proximity-induced *f*-*p* coupling—is a key guiding principle for designing novel magnetic TSs suitable for external control of topological phases and transport. This proposed model can aid the search for promising magnetic TSs and help pave the way towards realistic applications for spin-driven topological electronics [49].

## ACKNOWLEDGEMENTS

The authors thank H. Masuda, Y. Ishida, K. Akiba, A. Miyake, M. Tokunaga, T. Suzuki, T. Kurumaji, J. G. Checkelsky, T. Nomoto, M. Hirayama, R. Arita, and M. Kawasaki for fruitful discussions. We also acknowledge T. Akiba for his assistance with measurements. This study was supported in part by KAKENHI (Grant No. 17H01195, 17H06137, 19H02424, and 19K14652), JST PRESTO Hyper-nano-space Design toward Innovative Functionality (Grant No. JPMJPR1412), JST CREST (Grant No. JPMJCR16F1), and the Asahi Glass Foundation. This study was designed by AHM, HT, and SI. Single crystals were grown by AHM with inputs from AN. AHM and HT performed the structural, magnetic, and transport measurements. The theoretical calculations were done by MSB. AN and HS contributed to the discussion of the results. AHM, MSB, and SI wrote the manuscript with feedback from all the authors.

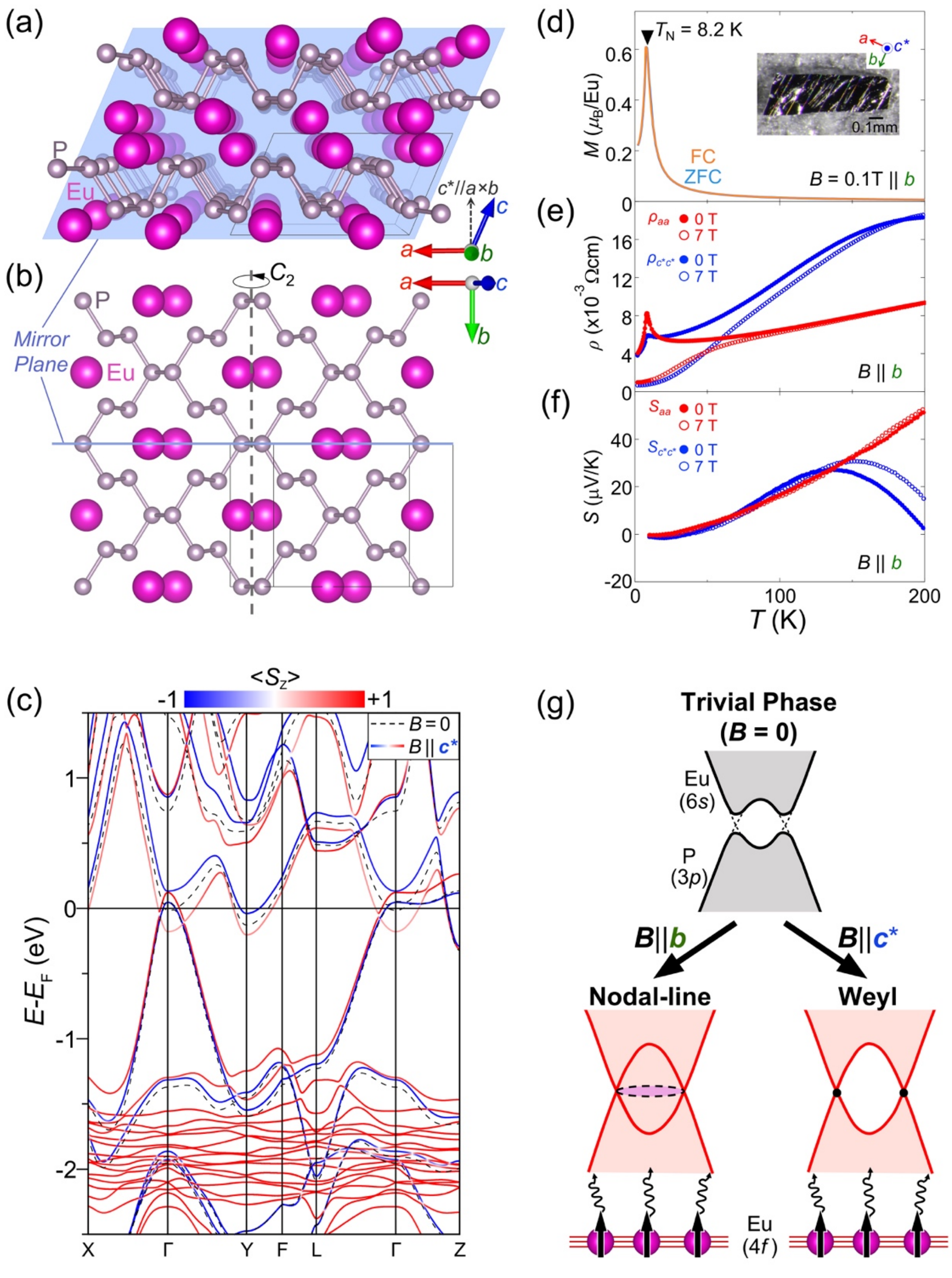

FIG. 1. Proximity-induced *f*-*p* exchange coupling in α-$EuP_3$. Side (a) and top (b) views of the monoclinic crystal structure of α-$EuP_3$ with unique axis *b*. The crystal has a single mirror plane perpendicular to the unique axis. The axis *c** is depicted, along with crystalline axes *a*, *b*, *c*, where we define ***a****, ***b****, ***c**** as ***a**** ∥ ***b*** × ***c***, ***b**** ∥ ***c*** × ***a*** (∥ ***b***), and ***c**** ∥ ***a*** × ***b***, respectively. (c) The electronic structure of α-$EuP_3$ in the paramagnetic ($B = 0$, dashed lines) and field-induced ferromagnetic (colored lines) phases. The localized 4*f*-bands lie approximately 1.5 eV below the Fermi level. In the FM phase calculation, $Eu^{2+}$ *f*-moments are fully aligned along the *c**-axis, corresponding to $M = 7\ \mu_B$/Eu, and the color of the bands represents the spin polarization.

(d) and (e) Temperature dependences of magnetization *M* and resistivity *ρ*, respectively, through the antiferromagnetic transition temperature $T_N$ = 8.2 K. *M* has been measured under field cooling (FC) and zero-field cooling (ZFC) conditions. A picture of the sample is given in the inset. (f) Temperature dependence of the Seebeck coefficient *S* down to 10 K. (g) Schematic picture of the band topology engineering in α-$EuP_3$. Nodal-line semimetal (***B*** || ***b***) and Weyl semimetal (***B*** || ***c****) phases are switched by the direction of *B* and the non-trivial band crossings can be tuned to the Fermi level via proximity-induced *f*-*p* exchange coupling.

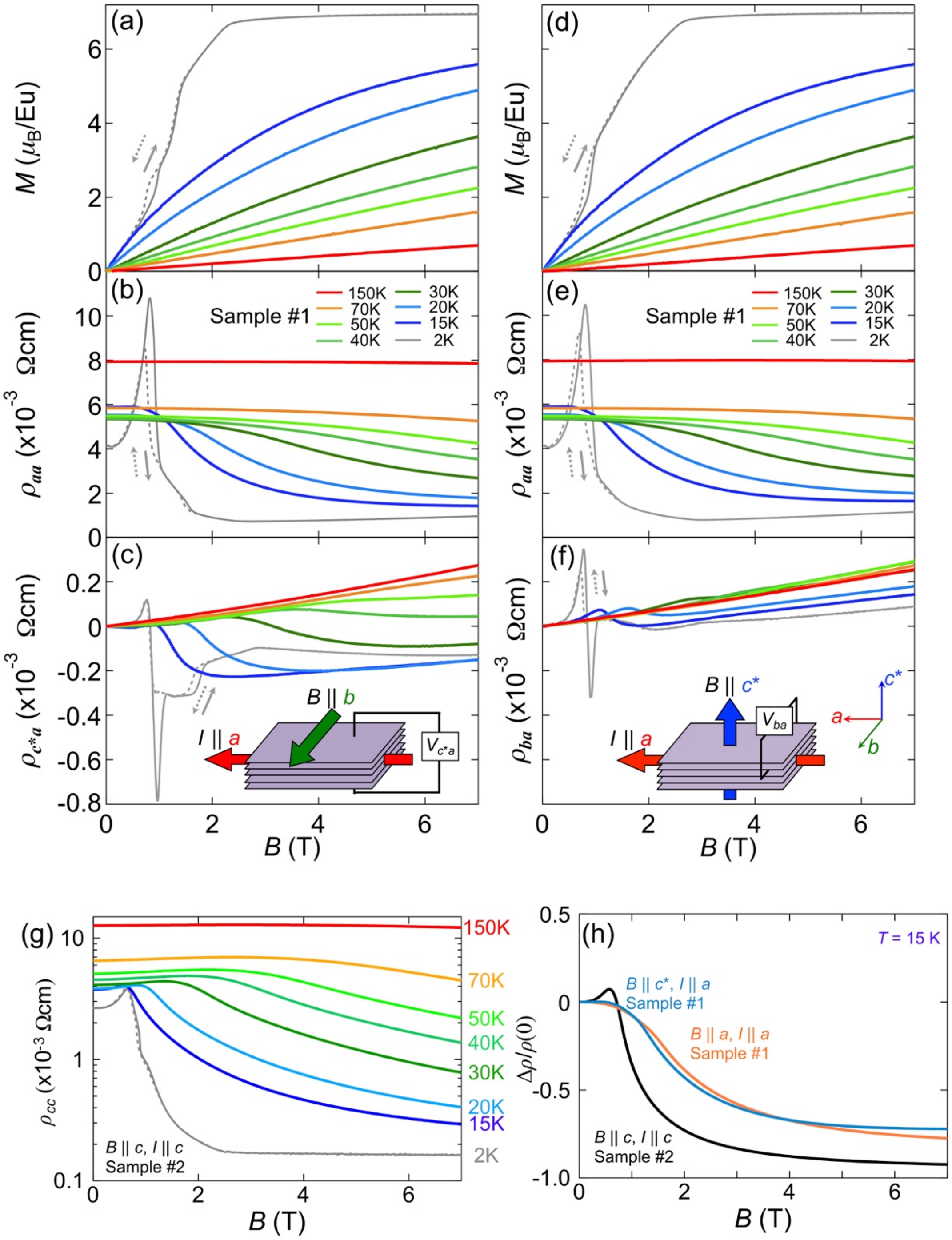

FIG. 2. Magnetic and transport properties of α-$EuP_3$. Magnetization, longitudinal resistivity, and Hall resistivity as functions of $B$ along the $b$-axis (a, b, and c) and $c$*-axis (d, e, and f). The insets show the schematics of the measurement configuration. (g) $\rho_{cc}$ measured with $\boldsymbol{I} \parallel \boldsymbol{B} \parallel \boldsymbol{c}$. (h) Magnetoresistance $\Delta\rho/\rho(0) = \{\rho(B) - \rho(0)\}/\rho(0)$ measured in three configurations at 15 K.

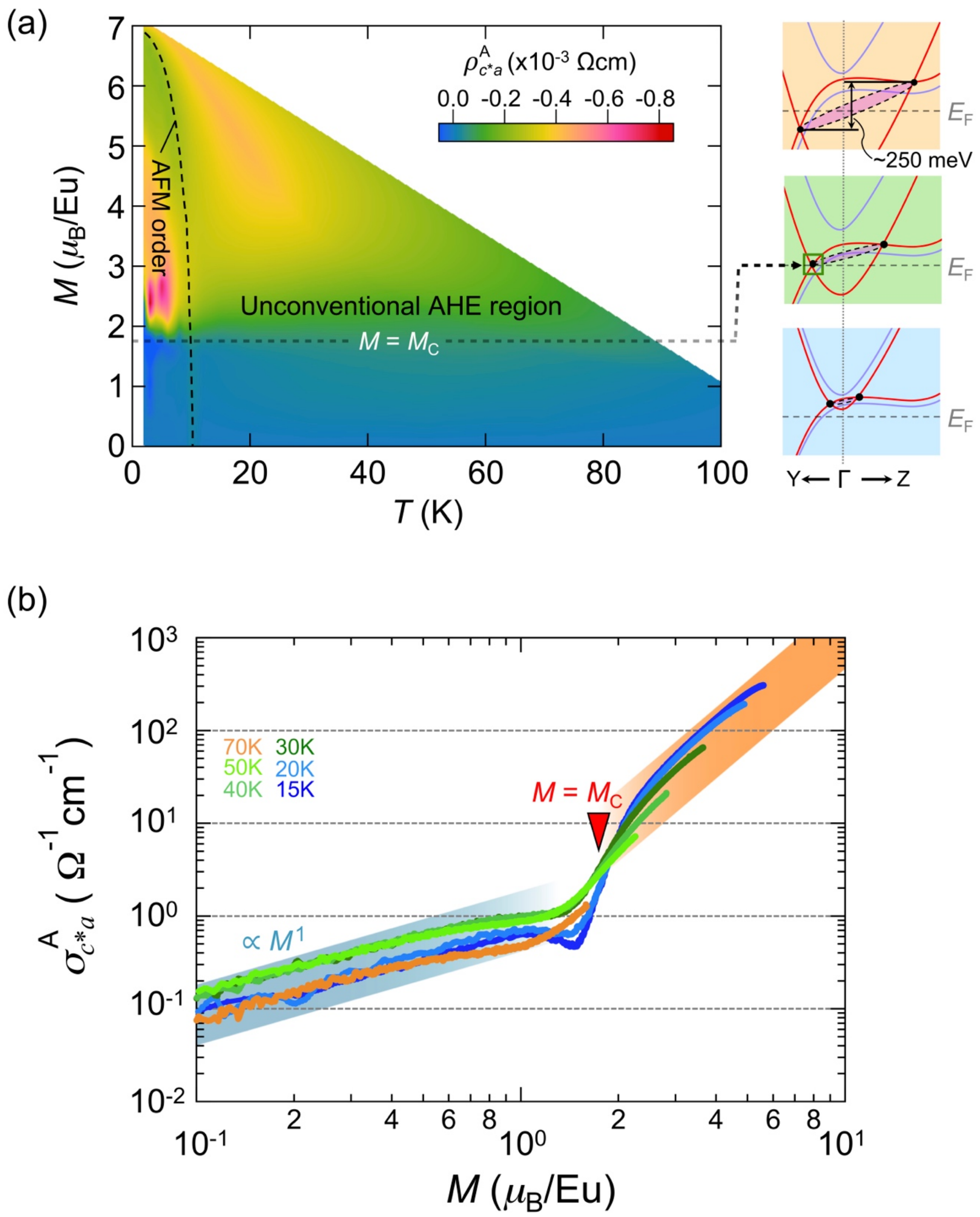

FIG. 3. Anomalous Hall effect in α-$EuP_3$ under $\boldsymbol{B} \parallel \boldsymbol{b}$. (a) Color plot of the anomalous Hall resistivity $\rho_{c*a}{}^{A} = \rho_{c*a} - \rho_{c*a}{}^{T = 150\ K}$ as a function of $M$ and $T$. 'AFM' denotes 'antiferromagnetic.' The schematics on the right side illustrate the $M$-dependent evolution of the nodal-ring shown in Fig. 4(d). Further explanation based on DFT calculations are presented in the Supplemental Material [37]. (b) Anomalous Hall conductivity $\sigma_{c*a}{}^{A} = -\rho_{c*a}{}^{A}/(\rho_{aa}\rho_{c*c*} - \rho_{ac*}\rho_{c*a})$ as a function of $M$. Further discussions on the analysis are presented in the Supplemental Material [37].

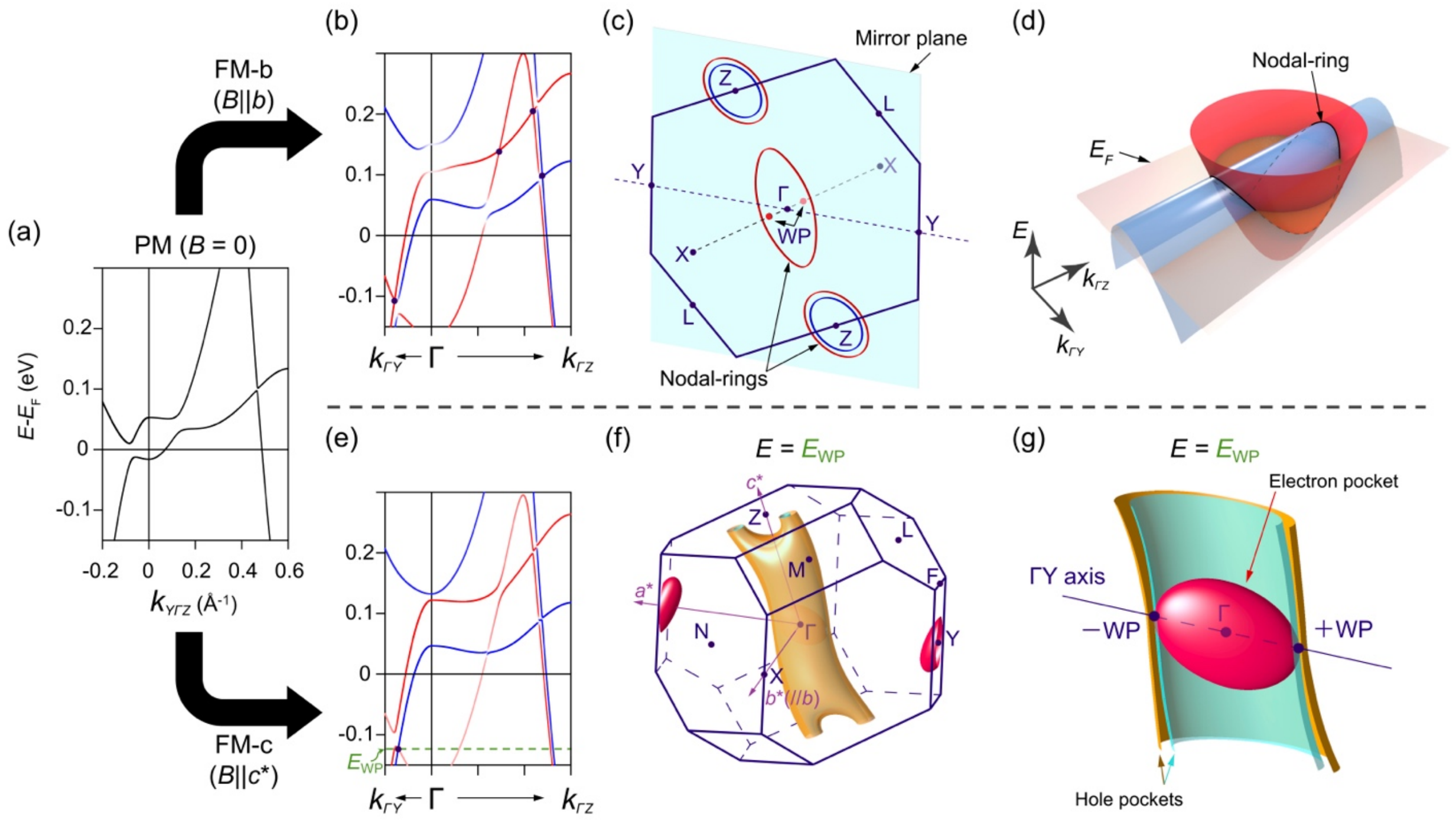

FIG. 4. Electronic structure of α-$EuP_3$ in the paramagnetic and field-induced ferromagnetic phases in different field directions. Energy dispersions of α-$EuP_3$ along Y-Γ-Z in the paramagnetic (PM) phase at $B$ = 0 (a), and field-induced ferromagnetic phases corresponding to $\boldsymbol{B} \parallel \boldsymbol{b}$ (FM-b) (b) and $\boldsymbol{B} \parallel \boldsymbol{c}^*$ (FM-c) (e). $E_{WP}$ in panel (e) denotes the energy level of the Weyl point. The dots in (b) and (e) show the symmetry protected band crossings. (c) Schematic illustration of topological nodal lines and Weyl nodes emerging in the vicinity of $E_F$ in phase FM-b. The nodal-rings emerge within the mirror plane and the Weyl nodes along Γ-X (see Supplemental Material [37]). (d) Schematic illustration of the nodal-ring around the Γ-point in phase FM-b. The area encompassed by the nodal-ring is not flat in energy and spans across $E_F$ at large $M$. (f) Brillouin zone and constant energy surface plot at $E = E_{WP}$ in phase FM-c. The electronic structure is composed of cylinder-like hole pockets stretched along the Γ-Z direction, reflecting the quasi-2D nature of this system, plus small electron pockets centered at the Γ- and Y-points. (g) Electronic structure around the Γ-point at $E = E_{WP}$ in phase FM-c, showing the presence of a single Weyl node pair along Γ-Y.

TABLE I. Magnitudes of the anomalous Hall angle ($|\Theta_{AHE}| = |\tan^{-1}(\sigma_H^A/\sigma)|$) and measurement temperatures of various AHE materials.

| Material | $\|\Theta_{AHE}\|$ (°) | $T$ (K) | Ref. |
|---|---|---|---|
| α-$EuP_3$ | 16.04-23.74 | 15 | This work [37] |
| $Co_{3-x}Fe_xSn_2S_2$ | 11.30-18.26 | 10-120 | [6],[38] |
| GdPtBi | 9.09-17.74 | 2.5 | [4] |
| $Co_2MnAl$ | 11.85 | 300 | [39] |
| $Fe_{3-x}GeTe_2$ | 2.40-4.91 | 2-10 | [7] |
| $KV_3Sb_5$ | 1.03 | 2 | [40] |
| MnGe | 0.68 | 5 | [41] |
| $Nd_2Mo_2O_7$ | 0.63 | 2 | [42] |
| Fe film | 0.63 | 300 | [43] |